\documentclass[conference]{IEEEtran}
\IEEEoverridecommandlockouts
\usepackage{cite}
\usepackage{amsmath,amssymb,amsfonts}
\usepackage{algorithmic}
\usepackage{graphicx}
\usepackage{subfig}
\usepackage{textcomp}
\usepackage{xcolor}
\usepackage{listings}
\usepackage[frozencache=true,cachedir=.]{minted}
\def\BibTeX{{\rm B\kern-.05em{\sc i\kern-.025em b}\kern-.08em
    T\kern-.1667em\lower.7ex\hbox{E}\kern-.125emX}}
\begin{document}

\title{BiRank vs PageRank: Using SNA on Company Register Data for Fiscal Risk Prediction
\thanks{This work is funded by the Austrian Federal Ministry of Finance as part of the Predictive Multigraph Analytics project.}
}

\author{\IEEEauthorblockN{1\textsuperscript{st} G\"oschlberger Bernhard} 
\IEEEauthorblockA{\textit{Research Studio Data Science} \\
\textit{Research Studios Austria FG}\\
Vienna, Austria \\
bernhard.goeschlberger@researchstudio.at}
\and
\IEEEauthorblockN{2\textsuperscript{nd} Deliu Dragos}
\IEEEauthorblockA{\textit{Predictive Analytics Competence Center} \\
\textit{Federal Ministry of Finance}\\
Vienna, Austria \\
dragos.deliu@bmf.gv.at}
}

\maketitle

\begin{abstract}
Efficient financial administrations need to ensure compliant behavior of all tax subjects without excessive personnel costs or obstruction of compliant companies.
To do so, accurate prediction of non-compliance or fraud is crucial.
Social Network Analysis (SNA) provides powerful tools for fraud prediction as fraudulence is often clustered in certain areas of real world social networks.
In this paper we present our results of comparing PageRank and the more recent BiRank to infer risk-ranks based on network structure and prior fraud information.
Specifically, we model our social network from company register data.
We find that in this case study BiRank outperforms PageRank in both quality of the resulting ranks for fraud prediction and run time.
The results show that this class of algorithms is generally useful for fraud and risk prediction and more specifically also illustrate the potential of BiRank in comparison, as it opens up new modeling opportunities.
Our results show that selecting companies for tax audits based on BiRank yields a precision of $16.38\%$ for the top 20.000 subjects selecting $83.4\%$ of all fraud cases (recall).
\end{abstract}


\section{Introduction}

Fiscal administrations are interested in compliant behavior of all tax subjects.
Beyond the bare financial aspect, a main motivation to enforce legal obligations such as taxes, duties and fees, is to maintain fair competition and a level playing field for all market entities.
However, a high density of tax audits results in an enormous bureaucratic overhead for both sides, companies and the administration.
An accurate prediction of non-compliant or even fraudulent behavior is therefore seen as a key challenge for administrations around the world.

In this paper we present preliminary results of a case study at the Austrian Federal Ministry of Finance (BMF).
To maximize the precision and outcome of audits, the BMF uses a wide range of approaches for the prediction of non-compliant behavior as a basis for audit selection.
The presented case study explores novel graph based influence models or spreading models to leverage social network information for fraud prediction.
More specifically, we implemented personalized BiRank\cite{He16} for a bipartite social network based on company register data to rank companies based on known risk-information.
We compare the approach against PageRank on a unipartite graph based on the same data.

\section{Research Question and Objectives}

This work is motivated by a practical problem: predicting fraudulent or non-compliant tax subjects based on previous audit information and publicly available information.
Our research objectives are therefore
\begin{enumerate}
    \item to identify data sources that provide information about real-world social networks and verify that these networks have a systematic effect on the distribution of fraud and non-compliant behavior among tax subjects
    \item to improve precision and recall of a top-n tax audit selection (with $n$ being a fixed amount of selected cases)
\end{enumerate}
As for research objective 1, we identified the Austrian company register as a publicly available data source of which we expect that it represents business related social relations.
As for research objective 2, we compare the results of PageRank and BiRank in terms of rank correlation, effect significance, and implications for tax audit selection precision and recall.   
We can state two underlying research questions:
\begin{enumerate}[\IEEEsetlabelwidth{RQ 1:}]
    \renewcommand{\labelenumi}{RQ \arabic{enumi}:}
    \item Is the company register's social network information a good predictor for non-compliant behavior or fraud?
    \item Which algorithm and modeling approach is best suited in terms of quality of ranks for prediction and run time?
\end{enumerate}

\section{Related Work}

Fiscal risk assessment uses fraud prediction models to evaluate risk of non-compliance of tax paying entities.
Social Network Analysis (SNA) provides an additional, very specific angle to deduct information for risk prediction (or fraud prediction). SNA is a technique in which entities are represented as nodes and their relationship as links and different graph analysis techniques can be used to identify, among others, suspicious entities, relationships or anomalies in the network.

A wide range of SNA methods that are (or can be) used for fraud prediction exists in the literature. Network visualization and ego-centric analysis (and quite similarly the snowball method) can be used to identify links with known fraudulent entities (for example in~\cite{Kodate2020}).
Collective inference or collective classification methods, such as Gibbs sampling or loopy believe propagation, are commonly used in~\cite{baesens2015fraud}.
The GOTCHA! propagation algorithm diffuses fraud through the network, labeling the unknown and anticipating future fraud while simultaneously decaying the importance of past fraud. 
This was used in~\cite{Gotcha} to identify companies that intentionally go bankrupt to avoid contributing their taxes.
Features of network topology (e.g. stars, cliques, string or grid-based structures) can be analyzed such that structure that is inconsistent with the rest can be identified and anomalies (outliers) in the network are detected~\cite{Dao2021}.

The PageRank algorithm, originally presented as a method to rank websites~\cite{page99}, is a widely known iterative collective inference method, inspired by eigenvectors.
Contrasting eigenvectors, PageRank calculation can be biased using a personalization or (re)start vector.
While designed for personalized web search engines originally, it can also be applied to fraud prediction.
In this case, known fraud information is used in the personalization vector.

BiRank, which was introduced in~\cite{He16}, ranks vertices of a bipartite graph (a graph whose vertices can be divided into two disjoint and independent sets and such that every edge connects a vertex from one set to the other one). This method, similarly to other ranking techniques on bipartite graphs, ranks by iteratively propagating scores on the vertices of the graph, which eventually converge towards a stationary ranking.

Similarly to the damping factor in PageRank, BiRank uses two hyperparameters to combine the score calculated from the graph structure and query vector (with lower values of the parameters achieving quicker convergence than higher values). BiRank, unlike other ranking methods~\cite{Bias}, is not biased towards higher degree vertices.

One can further generalize this method to n-partite graphs and an application for TriRanks can be found in~\cite{trirank}.

\section{Austrian Company Register Data}

The Austrian Company Register (formerly the Commercial Register) contains entries in accordance with company law, such as disclosure of financial statements, persons authorized to represent the company or shareholder data.
It is a public directory maintained in a database by the courts of the Austrian federal provinces.
A large number of entries in the company register can be made electronically by entrepreneurs.
Anyone can inspect the company register.
Company information contained in the the company register is also available on the Internet, with small fees being charged for each request.
Also a number of services exist that provide similar, non-authoritative information for free\footnote{e.g. firmenabc.at}.

For our case study we use information about company representatives to construct social network graphs (as described in section \ref{subsec:nw-model-pr} and \ref{subsec:nw-model-br}).
The data on company representatives does contain information on their function, start date and end date (optionally).
For our case study we were specifically looking at managing directors and shareholder-managing directors and included 30 years of historical data from 1991 to 2021.

The company register has its own identifier for companies, the company register number.
This number is typically not used by the fiscal administration, as other idenfifiers exist for taxes, fees and duties.
Also the company register does not contain unique identifiers for persons which are only given by name and address.
Thus, the data can't be mapped onto identifiers used by the fiscal administration directly.
To harness the data of the company register, textual information, such as given name, surname, company name, address information, etc. that are present in both---Company Register and data sets of the fiscal administration---are used to resolve entities.
Generally, the data quality after entity resolution is very good, as almost all entities can be resolved and the resolved entities are accurate.
Our work builds upon this preprocessed company register data, and we will subsequently describe our network modeling approach.

\section{Graph Modeling Approaches}

A comparison of PageRank and BiRank needs to account for their differences in design.
While PageRank operates on unipartite graphs, BiRank is designed for bipartite graphs.
For our case study, only companies are subject of interest and need to be ranked.
Therefore, we use the same underlying data to model two different networks:
\begin{enumerate}
    \item a unipartite company-company network for PageRank
    \item a bipartite company-person network for BiRank
\end{enumerate}

In the following sections, we start by describing general modeling considerations, and subsequently the detailed construction of the unipartite and bipartite network.

\subsection{General Modeling Consideration}

Generally, both PageRank and BiRank, are designed to run on weighted directed graphs.
Edge weights and edge directions are a primary consideration for modeling the social network graph.
For the idea of information spreading/influence on the company register network case we decided to work on an undirected graph.

A possible consideration to model edge directions is the use of temporal information (e.g. start date: managing director moved from company A to company B).
However, this appeared impractical for multiple reasons:
\begin{enumerate}
    \item Many connections result from directors being active in multiple companies concurrently anyways.
    \item Companies are audited for the last seven years and non-compliant behavior might be found in companies where the managing director was active before the known non-compliant case.
    \item We want to use temporal information for time sliced modeling in future work. 
\end{enumerate}

For edge weights we used temporal information.
An edge weight between a company and a person is set such, that it is equal to the time a person has held the position (managing director or shareholder-managing director in years (rounded up).
Consequently, the weights range between 1 and 30.

Notably, there are two special cases in company register data that had to be handled specifically:
\begin{enumerate}
    \item One-person enterprises: Name and address of shareholder-managing director and company are identical. In this case entity resolution maps both to the same identifier.
    \item Managing firms: It is possible that a company serves as a managing director of another company under certain circumstances.
\end{enumerate}

Both cases would conflict with the bipartite property of the modeling approach for BiRank.
Consequently, we split them into two surrogate nodes and generate an artificial edge that connects them.
We set an arbitrary edge weight of 30 for these edges---which is the maximum edge weight in the network---as we consider them strongly connected, since they are surrogate nodes for the same entity, and prefer that they are assigned to the same partition. 

Finally, we partitioned the resulting graphs, and filtered out small partitions, to improve performance in terms of run time and memory footprint (see subsection~\ref{subsec:partitioning}).

\subsection{Network Modeling for PageRank}
\label{subsec:nw-model-pr}

Subject of our risk model are companies.
Consequently, entities that connect them, but are not subject to our risk prediction, can be modeled as edges.
For PageRank, we construct such a company-company network from the Company Register data.
We find three cases of company-company connections:
\begin{itemize}
    \item exactly one natural person serving as managing director of exactly two companies,
    \item more than one natural person serving (together) as managing directors of exactly two companies (multigraph), and
    \item exactly one natural person serving as managing director of three or more companies (hypergraph).
\end{itemize}

For the second case (multigraph), we sum up the edge weights of all edges to reduce it to a graph.
Consequently, multigraph edges reduce the number of edges in the company-company network compared to the bipartite company-person network.
For the third case (hypergraph), we create a fully connected graph for the nodes of the hyperedge.
Compared to the bipartite case where hyperedges are represented by $n+1$ edges---with $n$ being the number of companies---they result in $\frac{n*(n-1)}{2}$ for the company-company network.

The total network before filtering for PageRank consisted of $302995$ nodes and $4506002$ edges.

\subsection{Network Modeling for BiRank}
\label{subsec:nw-model-br}

The network for BiRank is a more direct representation of the company register information.
It contains nodes of two types: companies and (natural) persons.
As stated above, the exceptions of a one-person company and a managing firm have to be handled.
They are resolved by creating a seperate node in each partition (a company node and a person node).

The total network for BiRank before filtering consisted of $431227$ company nodes, $520745$ person nodes and $959693$ edges.

\subsection{Components and Spectral Partitioning}
\label{subsec:partitioning}

The network consists of 39787 connected components of which only 3 components contain more than 50 company nodes.
The largest component contains 190273 company nodes, and the second largest component contains only 61 company nodes.
For practical reasons, the largest component is further partitioned and we run the algorithms seperately on each partition to minimize effective runtime\footnote{runtime complexity remains unchanged} and memory footprint.
For partitioning, we recursively apply spectral bisection until the largest remaining partition is smaller than a threshold of 50000 company nodes.
Spectral bisection is based on the Fiedler vector---the second smallest eigenvector of the graph Laplacian---and splits along the signs of the values of the Fiedler vector.
The threshold was chosen as a trade-off at a point where the distribution of resulting partition sizes and expected run time appeared acceptable in practice.
To get comparable results the partition information is only calculated on the unipartite (company-company) network and used for both cases.
Partitions with less than 50 company nodes are ignored for two reasons: run time considerations and expected quality of results.
After this step 71 partitions consisting of 189630 company nodes remain.

\section{Implementation}

We implemented the algorithms in SAS/IML and encapsulated the implementation with the SAS Macro facility.
The macros also include shared logic, such as by-cluster processing, cross-validation and parallelization, that are identical for both algorithms.

We use 10-fold cross-validation for all experiments throughout this paper.
For 10-fold cross-validation each of the 10 folds contains only $\frac{9}{10}$ of the fraud information and the other $\frac{1}{10}$ is set to unknown.
The rank value (PageRank or BiRank) of a node is then always taken from the fold that does not contain its own risk information.
Consequently, the resulting rank values are solely based on risk values of other nodes and the network information.

The restart vectors for both algorithms are personalized with risk information.
For our experiment we used a binary risk indicator for tax subjects based on audit data and statistical models.
For subjects where this indicator is unknown, we use zero. 

\subsection{PageRank}

For the sake of completeness we list the core of our PageRank implementation in listing~\ref{lst:pageRank}.
It follows the traditional iterative algorithm, where $A$ is a row-normalized adjacency matrix, $e$ is the personalization (in our case risk) vector.
The algorithm terminates when convergence is reached or an iteration threshold is reached.

\begin{listing}[ht]
\begin{minted}{sas}
do while (i<max_iter & error>epsilon);
  r=r_;
  r_=alpha*
    SparseMulVecSym(A, r) + (1-alpha)*e;
  error = norm(r-r_, "L2");
  i = i + 1;
end;
\end{minted}
\caption{Iterative calculation of PageRank}
\label{lst:pageRank}
\end{listing}

\subsection{BiRank}

The iteration logic of our BiRank implementation is shown in listing~\ref{lst:biRank}, and is very similar to the PageRank implementation.
The main difference is the use of the normalized transition matrices $S$ (and the transposed transition matrix $S_T$), which uses the inverse of the product of the square roots of the node degrees to dampen transition weights (i.e. $S=D_{u}^{-1/2} W D_{p}^{-1/2}$, respectively $S_{ij}=\frac{W_{ij}}{\sqrt{d_i}*\sqrt{d_j}}$). 

\begin{listing}[ht]
\begin{minted}{sas}
do while (i<max_iter & error>epsilon);
  u_ = alpha*
    SparseMulVecSym(S_T, p)+(1-alpha)*u0;
  p_ = beta*
    SparseMulVecSym(S, u_)+(1-beta)*p0;
  error = sum(abs(p-p_)) + sum(abs(u-u_));
  p = p_; u = u_;
  i = i + 1;
end;
\end{minted}
\caption{Iterative calculation of BiRank}
\label{lst:biRank}
\end{listing}

\section{Performance}

While we did not run extensive run time tests under laboratory conditions and on different graphs, we found BiRank to outperform PageRank substantially for our use case.

Notably, the bipartite modeling approach resulted in a larger graph in terms of nodes ($951972 > 302995$), but was actually considerably smaller in terms of edges ($959693 < 4506002$).
Hence, the unipartite network was also more dense.

Our measurements are based on 20 runs.
A BiRank run over all partitions took $112.45 min$ on average, while a PageRank run over the same amount of partitions took $258.9 min$ on average.
The run time on our system was relatively stable with a standard deviation of $5.32 min$ for BiRank and $1.87 min$ for PageRank.

\section{Rank Quality}

As the names of both algorithms suggest, the results of PageRank and BiRank should be treated as ordinal data.
Therefore, we use Spearman's rank correlation coefficient and the Mann-Whitney U test~\cite{mann47} (also known as Wilcoxon rank-sum test).

Table~\ref{tab:spearman} shows the correlation of the calculated rank values with the binary risk indicator.
Both cases show a highly significant positive correlation.
Since the rank values are calculated with cross-validation, they do not contain the original risk indicator information for the node they were calculated for.
The results therefore indicate that the risk information about the node itself (ground truth) and the risk information about other nodes that \textit{influence} the node (dependent on the network structure) are positively correlated.

\begin{table}[hbt]
    \centering
    \renewcommand{\arraystretch}{1.3}
    \caption{Spearman's $\rho$ for the rank correlation of the calculated ranks and the binary risk indicator. $N$ is the total number of tax subjects in the network, $n_0$ is the number of tax subjects that are known to be compliant, and $n_1$ is the number of non-compliant or fraudulent subjects.}
    \label{tab:spearman}
    \begin{tabular}{|*{6}{c|}}
        \hline
        & $\rho$ & $p$ & $N$ & $n_0$ & $n_1$ \\ \hline
        BiRank & $0.24893$ & $<.0001$ & $189530$ & $92509$ & $3928$ \\
        PageRank & $0.14646$ & $<.0001$ & $174771$ & $84626$ & $3617$ \\  \hline
    \end{tabular}
\end{table}

\begin{figure*}[!t]
\centering
\subfloat{
    \includegraphics[width=3.2in]{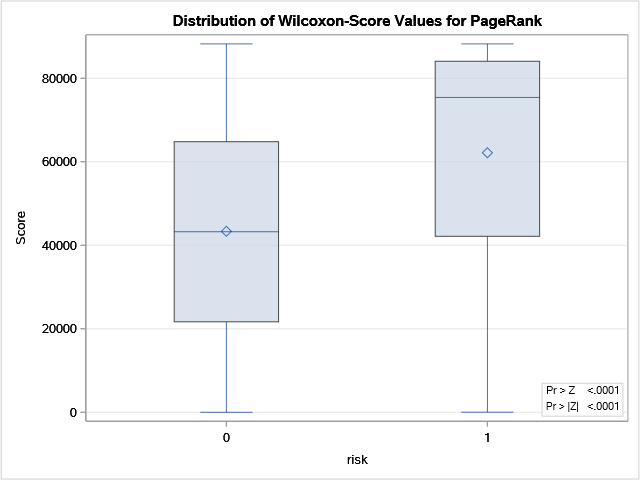}
    }
\hfil\subfloat{
    \includegraphics[width=3.2in]{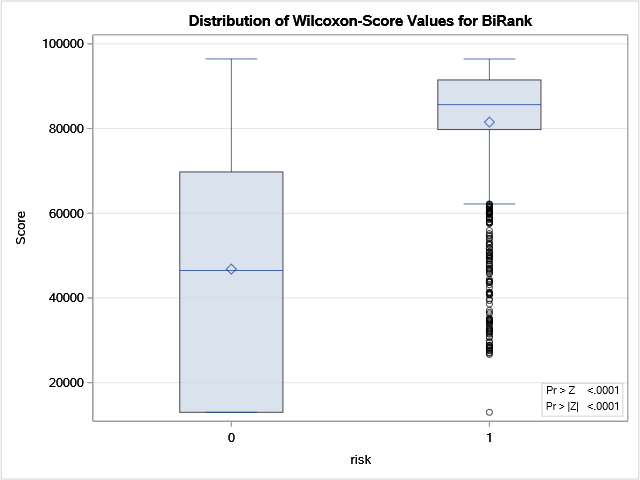}
    }
\caption{Boxplot of the Wilcoxon-Ranks for PageRank (left) and BiRank (right) by the binary risk indicator.}
\label{fig:boxplots}
\end{figure*}

We applied the Mann–Whitney U test and found both PageRank ($Z=42.48$) and BiRank ($Z=71.04$) to yield highly significant results ($p < 0.001$), showing that the calculated PageRank and BiRank are significantly higher for subjects with risk indicator $= 1$.
Figure~\ref{fig:boxplots} shows the distribution of the wilcoxon ranks for PageRank and BiRank.
It illustrates the stronger effect between the two groups for the BiRank results, as the distributions appear less overlapping. 

\begin{figure*}[!t]
\centering
\subfloat{
    \includegraphics[width=3.2in]{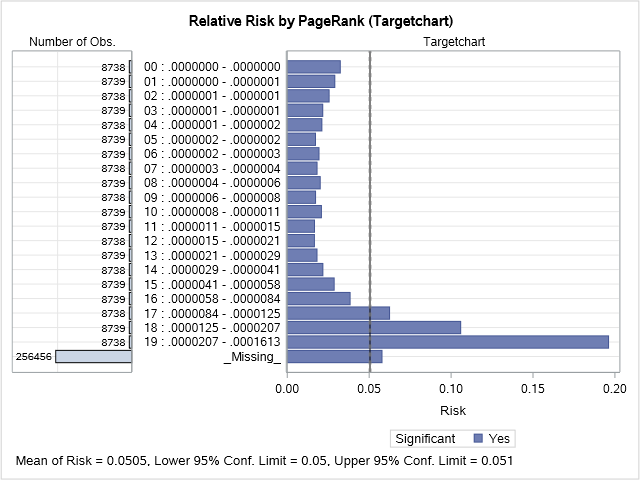}
    }
\hfil\subfloat{
    \includegraphics[width=3.2in]{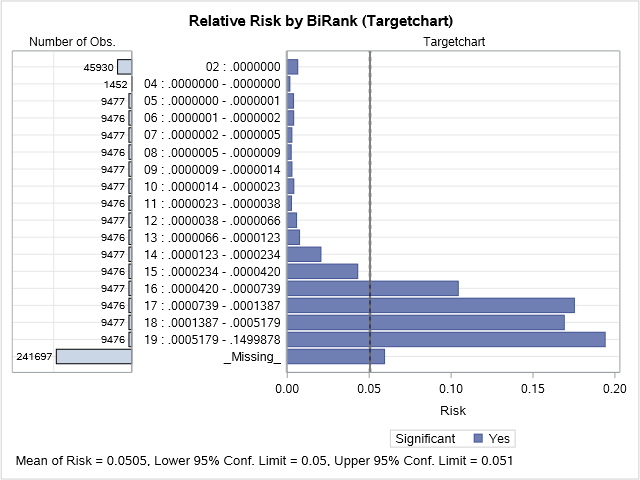}
    }
\caption{Target chart for PageRank (left) and BiRank (right): It splits the ranked data into equally sized bins and each bar shows the relative number of subjects with risk~indicator~=~1.}
\label{fig:target_charts}
\end{figure*}

To visually illustrate the value for risk prediction we also use a \textit{target chart}, a bar chart showing the relative amount of subjects with risk indicator 1, for equally sized, ordered bins.
Such \textit{target charts} for binary risk indicators show the lift that a value---in this case PageRank and BiRank results---can provide for predictions.   
The \textit{target charts} for our results are depicted in figure~\ref{fig:target_charts}.

\section{Discussion}

The results of PageRank and BiRank are significant regarding the risk indicator in the Wilcoxon rank-sum test.
As illustrated by the \textit{target charts} in figure~\ref{fig:target_charts}, the top ranked $5\%$ of both algorithms contain more than 7 times more subjects with a risk value of 1 than the overall population.
These results validate the use of SNA and collective inference methods for risk prediction (or fraud prediction). 

For our use case BiRank outperforms PageRank in terms of run time and quality of the resulting ranks.
Our work is not an extensive analysis of the algorithmic properties but a case study on a real world application.
It illustrates the impact of the network shape on suitability of modeling approaches.
The combinatorial creation of edges between hypergraph edges (one person connecting multiple companies) for the company-company network we used for PageRank resulted in a substantially denser network with about $4.5$ times more edges than in the bipartite network used for BiRank.
High degree person nodes (i.e. persons serving as managing director in a large number of companies) cause this increase in edges.
We found that the top 20 person nodes in terms of degree create $602421$ edges in the unipartite network, and that $40627$ persons have served as managing director in more than 3 companies over the last 30 years (leading to more edges in the unipartite network than in the bipartite network). 

\begin{figure}
    \centering
    \includegraphics[width=\linewidth]{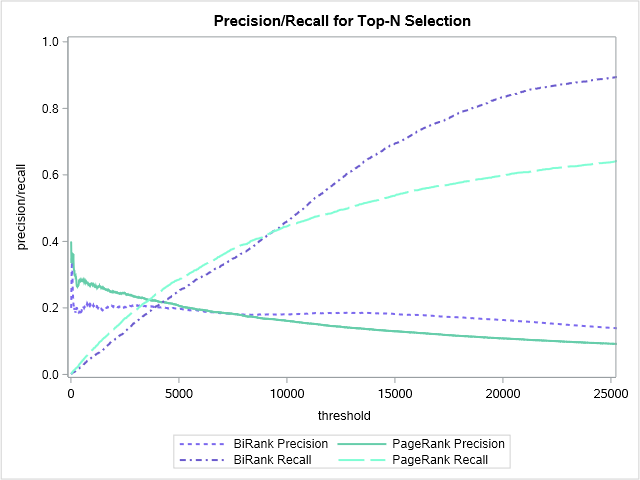}
    \caption{Precision and recall of top ranked tax subject selection.}
    \label{fig:precision-recall}
\end{figure}

In terms of result quality BiRank (and the corresponding modeling approach) produce more accurate results for tax audit selections with more than 7500 cases.
As illustrated by the rank distribution in figure~\ref{fig:boxplots} and the \textit{target charts} in figure~\ref{fig:target_charts} BiRank misses less subjects with risk indicator 1 than PageRank, which assigns low ranks to a higher proportion of risk subjects.
The \textit{target chart} illustrates this quite well, as we see a significant lift greater than 3 for the last three bins for BiRank and only in the last bin for PageRank.
In terms of precision PageRank performs better for the top-ranked subjects, peaking at $0.4$ for the top 10 cases and a precision of $>0.3$ for the top 160 cases.
Conversely, the precision for BiRank based selection is lower for small selections ($0.34$ for top 50), but significantly surpasses PageRank based selections in terms of precision when the number of selected cases is greater than 7500.
Figure~\ref{fig:precision-recall} illustrates the precision and recall for tax audit selection based on both rankings dependent on the number of selected cases for known fraud information.
Additionally, Table~\ref{tab:auditselection} shows selected values for these measures.
For a selection of 20.000 tax subjects based on BiRank ($20.7\%$ of all subjects with known information about tax compliance), we identify 3276 true-positives ($83.4\%$ of all non-compliant/fraudulent cases).

\begin{table}[hbt]
    \centering       
    \caption{Fraud cases in tax audit selections based on BiRank and PageRank.}
    \begin{tabular}{|*{5}{c|}}
        \hline
        \# of Audits & \multicolumn{2}{c|}{Precision} & \multicolumn{2}{c|}{Recall} \\
        & BiRank & PageRank & BiRank & PageRank \\ \hline
        100 &	23.00\% &	35.00\% &	0.59\% &    0.97\% \\
        1000 &	20.90\% &	27.10\% &	5.32\% &	7.49\% \\
        5000 &	19.78\% &	20.62\% &	25.18\% &	28.50\% \\
        10000 &	18.08\% &	16.11\% &	46.03\% &	44.54\% \\
        20000 &	16.38\% &	10.82\% &	83.40\% &	59.83\% \\ \hline
    \end{tabular}
    \label{tab:auditselection}
\end{table}

\section{Conclusion and Outlook}

We found that the general claims about speed of convergence and performance stated by \cite{He16} to be in line with our case study findings.
Based on our results, we argue that we can answer RQ~1, as we find the company register based social network to be a good predictor for fraud or non-compliance.
With respect to RQ~2, we find that for our case BiRank yields the best results overall.
The strong lift for highly ranked subjects in terms of risk validates the value of SNA methods for risk and fraud prediction in general, but more importantly it demonstrates that BiRank is particularly useful for cases like the presented one.

For future work we plan to experiment with time sliced versions of the graphs and to explore temporal effects on non-compliant or fraudulent behavior.
We expect that the increase in edges for the unipartite network declines for time sliced versions of the graph, as a shorter time window may result in a lower amount of persons acting as managing director or 3 or more companies, and a lower amount of companies that a single person is related to in general.
If this holds true, the performance of PageRank should increase.
However, we still expect BiRank to yield better results, as it seems to avoid the unwanted effect of overemphasizing central nodes and nodes that are directly connected to them.

We also expect to get better results by tuning edge weights.
Additional information on companies (e.g. turnover, industry sector) and persons (e.g. income and sources of income) can be used to construct more meaningful weights.

Another aspect for further exploration is the risk value used for the restart vector.
Depending on the information source used, the Ministry of Finance has a plethora of different risk values for tax subjects.
We used a binary indicator based on both audit data and statistical models for now, but many other options exist, such as values solely based on audit data or more differentiated values on a continuous scale.
For future work, we will explore for which type of risk indicator the network yields the best predictions.

We also plan to include shareholder information in future versions of the network.
As shareholders tend to be companies themselves very often, it remains open to see if our approach to maintain a bipartite graph is still suitable for that case.

Finally, we plan to analyze the impact of the calculated ranks on the multi-model that is used for final scoring of tax subjects, to see if our approach adds new information to the scoring model or if that information is already mostly covered through other models feeding into it.

\bibliographystyle{IEEEtran}
\bibliography{refs}

\end{document}